\documentclass[twocolumn,showpacs,aps,prl,superscriptaddress,letterpaper]{revtex4}

\usepackage{graphicx}
\usepackage{dcolumn}
\usepackage{amsmath}
\usepackage{epsfig}
\usepackage{multirow}

\long\def\inst#1{\par\nobreak\kern 4pt\nobreak
    {\itshape #1}\par\vskip 10pt plus 3pt minus 3pt}
\RequirePackage{xspace}
\usepackage{relsize}

\def\babar{\mbox{\slshape B\kern-0.1em{\smaller A}\kern-0.1em
    B\kern-0.1em{\smaller A\kern-0.2em R}}}
\def\Abar    {\kern 0.18em\overline{\kern -0.18em A}{}\xspace}
\def\Kbar    {\kern 0.18em\overline{\kern -0.18em K}{}\xspace}
\def\Bbar    {\kern 0.18em\overline{\kern -0.18em B}{}\xspace}
\def\Dbar    {\kern 0.18em\overline{\kern -0.18em D}{}\xspace}

\def\BB      {\ensuremath{B\Bbar}\xspace} 
\def\Bz      {\ensuremath{B^0}\xspace}
\def\Bzb     {\ensuremath{\Bbar^0}\xspace}
\def\BzBzb   {\ensuremath{\Bz {\kern -0.16em \Bzb}}\xspace}
\def\Bu      {\ensuremath{B^+}\xspace}
\def\Bub     {\ensuremath{B^-}\xspace}

\def\BpBm    {\ensuremath{\Bu {\kern -0.16em \Bub}}\xspace}

\newcommand{\optbar}[1]{\shortstack{{\tiny (\rule[.4ex]{1em}{.1mm})}
  \\ [-.7ex] $#1$}}
\def\BorBbar    {\kern 0.18em\optbar{\kern -0.18em B}{}\xspace}
\def\DorDbar    {\kern 0.18em\optbar{\kern -0.18em D}{}\xspace}
\def\KorKbar    {\kern 0.18em\optbar{\kern -0.18em K}{}\xspace}

\def\pep2{PEP-II}
\mathchardef\Upsilon="7107
\def\Y#1S{\ensuremath{\Upsilon{(#1S)}}\xspace}

\def\FourS {\Y4S}

\newcommand{\BABARPubYear}     {07}
\newcommand{\BABARPubNumber}  {023}

\newcommand{\SLACPubNumber} {12461}

\begin{document}

\begin{flushleft}
\babar-PUB-\BABARPubYear/\BABARPubNumber\\
SLAC-PUB-\SLACPubNumber
\\[10mm]
\end{flushleft}

\title{
\large \bfseries \boldmath
Search for $B^0\to\phi(K^+\pi^-)$ Decays with Large $K^+\pi^-$ Invariant Mass
}

%
\author{B.~Aubert}
\author{M.~Bona}
\author{D.~Boutigny}
\author{Y.~Karyotakis}
\author{J.~P.~Lees}
\author{V.~Poireau}
\author{X.~Prudent}
\author{V.~Tisserand}
\author{A.~Zghiche}
\affiliation{Laboratoire de Physique des Particules, IN2P3/CNRS et Universit\'e de Savoie, F-74941 Annecy-Le-Vieux, France }
\author{J.~Garra~Tico}
\author{E.~Grauges}
\affiliation{Universitat de Barcelona, Facultat de Fisica, Departament ECM, E-08028 Barcelona, Spain }
% ...
\collaboration{The \babar\ Collaboration}
\noaffiliation

\date{May 2, 2007}


\begin{abstract}
Motivated by the polarization anomaly in the $B\to\phi(1020)K^{*}(892)$
decay, we extend our search for other $K^*$ final states
in the decay $B^0\to\phi(1020)K^{*0}$ 
with the $K^{*0}\to K^+\pi^-$ invariant mass above 1.6 GeV.
The final states considered include the $K^{*}(1680)^{0}$, $K_3^{*}(1780)^{0}$, 
$K_4^{*}(2045)^{0}$, and a $K\pi$ spin-zero nonresonant component.
We also search for $B^0\to\phi\Dbar^0$ decay with the same final state.
The analysis is based on a sample of about
384 million $\BB$ pairs recorded with the $\babar$ detector.
We place upper limits on the branching fractions
${\cal B}(B^0 \to \phi K^{*}(1680)^{0})<3.5\times 10^{-6}$,
${\cal B}(B^0 \to \phi K_3^{*}(1780)^{0})<2.7\times 10^{-6}$,
${\cal B}(B^0 \to \phi K_4^{*}(2045)^{0})<15.3\times 10^{-6}$, 
and ${\cal B}(B^0 \to \phi \Dbar^0)<11.7\times 10^{-6}$
at 90\% C.L. The nonresonant contribution is consistent
with the measurements in the lower invariant mass range.
\end{abstract}


\maketitle


Recent measurements of polarization in rare vector-vector 
$B$ meson decays, such as $B\to\phi K^*$ and $\rho K^*$,
have revealed a large fraction of transverse 
polarization~\cite{babar:vv, belle:phikst, 
belle:rhokst, babar:rhokst, babar:vt}. 
This indicates a significant departure from the  
expected predominance of the longitudinal amplitude~\cite{bvv1}. 
The rate, polarization, and $C\!P$ measurements of $B$ meson decays
to particles with nonzero spin are sensitive to both strong and 
weak interaction dynamics and are discussed 
in a recent review~\cite{bvvreview2006,pdg2006}.

\begin{figure}[b]
\centerline{\setlength{\epsfxsize}{1.0\linewidth}\leavevmode\epsfbox{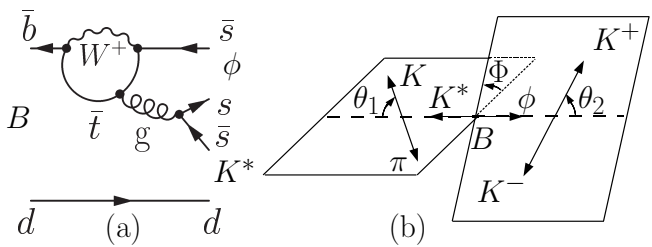}}
\caption{\label{fig:decay}
(a) Feynman diagram describing the $B^0\to\phi K^{*0}$ decay;
(b) definition of decay angles given in the rest frames of the decaying parents.
}
\end{figure}

In particular, the $B\to\phi K^*$ decays are 
potentially sensitive to physics beyond the 
standard model in the $b\to s$ penguin transition, 
shown in Fig.~\ref{fig:decay}~(a)~\cite{bvv1}.
The polarization anomaly in vector-vector $B$ meson decays
suggests other contributions to the decay amplitude, previously 
neglected. This has motivated a number of proposed 
contributions from physics beyond the standard model~\cite{nptheory}.
In addition, there are new mechanisms within the 
standard model which have been proposed to address the anomaly, 
including new weak dynamics~\cite{smtheory}, such as annihilation or 
electroweak penguin, or strong dynamics ~\cite{qcdtheory},
such as QCD rescattering. 

In order to distinguish the models,
the $\babar$ experiment extended the study 
of the $B^0\to\phi K^{*0}$ decays with the tensor ($J^P = 2^+$), 
vector ($J^P = 1^-$), and scalar ($J^P = 0^+$) $K^{*0}$~\cite{babar:vt}.
The vector-tensor results are in agreement with 
quark spin-flip suppression~\cite{bvv1} and $A_0$ amplitude dominance, 
whereas the vector-vector mode contains 
substantial $A_{+1}$ amplitude, corresponding to anomalously
large transverse polarization, where  
$A_{\lambda}$ corresponds to helicity 
states $\lambda=-1,0,+1$ of the $\phi$ and $K^{*}$ mesons.

In this paper, we extend our search for 
$B^0\to\phi K^{*0}$ to the higher-mass and higher-spin   
resonances $K^{*}(1680)^{0}$, $K_3^{*}(1780)^{0}$, 
and $K_4^{*}(2045)^{0}$. 
Charge conjugate reactions are implied throughout this paper.
The respective quantum numbers for these states 
$J^P = 1^-, 3^-,$ and $4^+$ are allowed in the
$K^{*0}\to K^+\pi^-$ decay. Moreover, we extend our study of the 
$B^0\to\phi(K\pi)_0^{*0}$ decay, where $(K\pi)_0^{*0}$ is 
the $J^P=0^+$ $K\pi$ component, to a $K\pi$ invariant mass
up to 2.15 GeV. We also search for the decay $B^0\to\phi\Dbar^{0}$,
which is expected to be significantly suppressed relative to the observed 
$B^0\to\omega\Dbar^{0}$ due to a negligible $u\bar{u}+d\bar{d}$ 
quark admixture in the $\phi$ meson~\cite{pdg2006}.

\begin{figure}[b]
\centerline{\setlength{\epsfxsize}{1.0\linewidth}\leavevmode\epsfbox{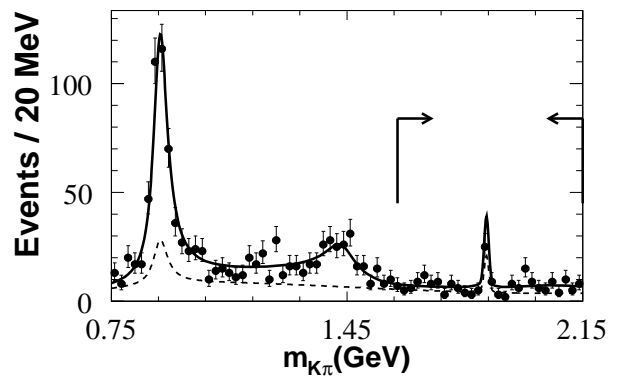}}
\caption{\label{fig:kpimass}
Distribution of the $K\pi$ invariant mass extended above 1.6 GeV
from the study of $B^0\to\phi(K^+\pi^-)$ decays in Ref.~\cite{babar:vt}.
The data distribution is shown with a requirement 
to enhance the signal as discussed in regard 
to Fig.~\ref{fig:projection1} in text.
The solid (dashed) line shows the signal-plus-background 
(background) expected distributions.
The arrows indicate the higher mass range,
1.60 to 2.15 GeV, used in this analysis.
}
\end{figure}

The analysis follows closely our recent study~\cite{babar:vt}
where we fully reconstruct the decay
$B^0\to\phi(1020) K^{*0}\to(K^+K^-)({K^+\pi^-})$.
The $K\pi$ invariant mass $m_{K\!\pi}$ window 
is now moved to the range from 1.60 to 2.15 GeV
to cover the above mentioned resonances. 
In Fig.~\ref{fig:kpimass} we show the $m_{K\!\pi}$ distribution 
extended from our previous study in Ref.~\cite{babar:vt}
to the mass range from 0.75 to 2.15 GeV.

\begingroup
\begin{table*}[t]
\caption
{\label{tab:results1}
Fit results for each decay mode:
the reconstruction efficiency $\varepsilon_{\rm long}$ 
and $\varepsilon_{\rm trans}$ obtained
from MC simulation for longitudinally and transversely
polarized events; 
the total efficiency $\varepsilon$, 
including the daughter branching fractions~\cite{pdg2006}
and assuming the smaller reconstruction efficiency; 
the number of signal events $n_{\rm sig}$; 
significance (${\cal S}$) of the signal; 
the branching fraction ${\cal B}$;
and the upper limit (UL) on the branching fraction at 90\% C.L.
The branching fraction ${\cal B}(B^0\to\phi (K\pi)^{*0}_0$) 
refers to the nonresonant $J^P=0^+$ $K\pi$ components
quoted for $1.60<m_{K\!\pi}<2.15$~GeV.
The systematic errors are quoted last
and are included in the ${\cal S}$ and UL calculations.
The negative event yield (or ${\cal B}$) 
for $B^0\to\phi K_3^{*}(1780)^0$ is extrapolated from
the likelihood distribution in the physical range.
}
\begin{center}
\begin{ruledtabular}
\setlength{\extrarowheight}{1.5pt}
\begin{tabular}{lccccccc}
\vspace{-3mm} & & & & & &  &\\
Mode  
 & $\varepsilon_{\rm long}$ (\%)  & $\varepsilon_{\rm trans}$ (\%) 
 & $\varepsilon$ (\%) & $n_{\rm sig}$ (events)
 & ${\cal S}$ ($\sigma$) & ${\cal B}$  ($10^{-6}$) & ${\cal B}$ UL ($10^{-6}$) \cr
\vspace{-3mm} & & & & & &  &\\
\hline
\vspace{-3mm} & & & & & &  &\\
 $\phi K^{*}(1680)^0$ 
 & $20.8\pm{2.9}$  & $21.6\pm{3.0}$ & $2.64\pm 0.41$  & $8^{+10}_{-7}\pm 11$ & $0.6$ & $0.7^{+1.0}_{-0.7}\pm 1.1$ & 3.5
\\
 $\phi K_3^{*}(1780)^0$ 
 & $27.7\pm{2.0}$   & $28.2\pm{2.1}$ & $1.71\pm 0.16$  & $-6\pm 10\pm 7$ & $0.0$ & $-0.9\pm{1.4}\pm 1.1$ & 2.7
\\
 $\phi K_4^{*}(2045)^0$ 
 & $23.6\pm{2.1}$   & $24.5\pm{2.2}$ & $0.77\pm 0.12$  & $18^{+14}_{-12}\pm 12$ & $1.2$ & $6.0^{+4.8}_{-4.0}\pm 4.1$ & 15.3
\\
 $\phi (K\pi)_0^{*0}$ 
 & $34.8\pm{1.6}$ & -- & $11.42\pm 0.56$  & $47\pm 16\pm 15$ & $2.2$ & $1.1\pm{0.4}\pm 0.3$ & 1.7
\\
 $\phi \Dbar^0$ 
 & $33.1\pm{1.6}$ & -- & $0.62\pm 0.03$  & $16\pm 7\pm 3$ & 2.4 & $6.5^{+3.1}_{-2.7}\pm{1.4}$ & 11.6
\\
\vspace{-3mm} & & & & & &  & \\
\end{tabular}
\end{ruledtabular}
\end{center}
\end{table*}
\endgroup

The angular distribution of the $B\to\phi K^*$ decay can be expressed 
as a function of ${\cal H}_i=\cos\theta_i$ and $\Phi$ shown in 
Fig.~\ref{fig:decay}~(b). Here $\theta_i$ with $i=1,2$
is the angle between the direction
of the $K$ meson from the $K^*\to K\pi$ ($\theta_1$) or $\phi\to K\Kbar$ 
($\theta_2$) and the direction opposite the $B$ in the $K^*$ or $\phi$
rest frame, and $\Phi$ is the angle between the decay planes of the two 
systems. For each decay mode,
the differential decay width has three complex amplitudes 
$A^J_{\lambda}$
corresponding to the spin of the $K\pi$ system $J\ge1$:
\begin{eqnarray}
\label{eq:helicityfull}
{d^3\Gamma \over d{\cal H}_1 d{\cal H}_2d\Phi} \propto
\left|~\sum_{\lambda=-1}^{+1} 
A^J_{\lambda} Y_{J}^{\lambda}({\cal H}_1,\Phi) Y_{1}^{-\!\lambda}(-{\cal H}_2,0)~\right|^2,
\end{eqnarray}
where $Y_{J}^{\lambda}$ are the spherical harmonics with 
$J=1$ for $K^{*}(1680)$,
$J=3$ for $K_3^{*}(1780)$, and 
$J=4$ for $K_4^{*}(2045)$.
The angular distribution is simplified when averaged over the
azimuthal angle $\Phi$ and becomes a function of the 
fraction of longitudinal polarization 
$f_L^J={|A^J_0|^2/(|A^J_{-1}|^2+|A^J_{0}|^2+|A^J_{+1}|^2)}$.
The angular distribution has only one contributing amplitude 
with $J=\lambda=0$ for each $\phi(K\pi)_0^{*}$ and $\phi\Dbar$ 
final state.


We use data collected with the \babar\ detector~\cite{babar} 
at the PEP-II $e^+e^-$ collider. A sample of $383.6\pm 4.2$ 
million $\FourS\to\BB$ events was recorded at the center-of-mass 
energy $\sqrt{s} = 10.58$ GeV. Charged-particle momenta are measured 
in a tracking system consisting of a silicon vertex tracker with five 
double-sided layers and a 40-layer drift chamber, both within the 1.5-T 
magnetic field of a solenoid. Charged-particle identification is provided 
by measurements of the energy loss in the tracking devices and by 
a ring-imaging Cherenkov detector.

\begin{figure*}[t]
\centerline{
\setlength{\epsfxsize}{0.35\linewidth}\leavevmode\epsfbox{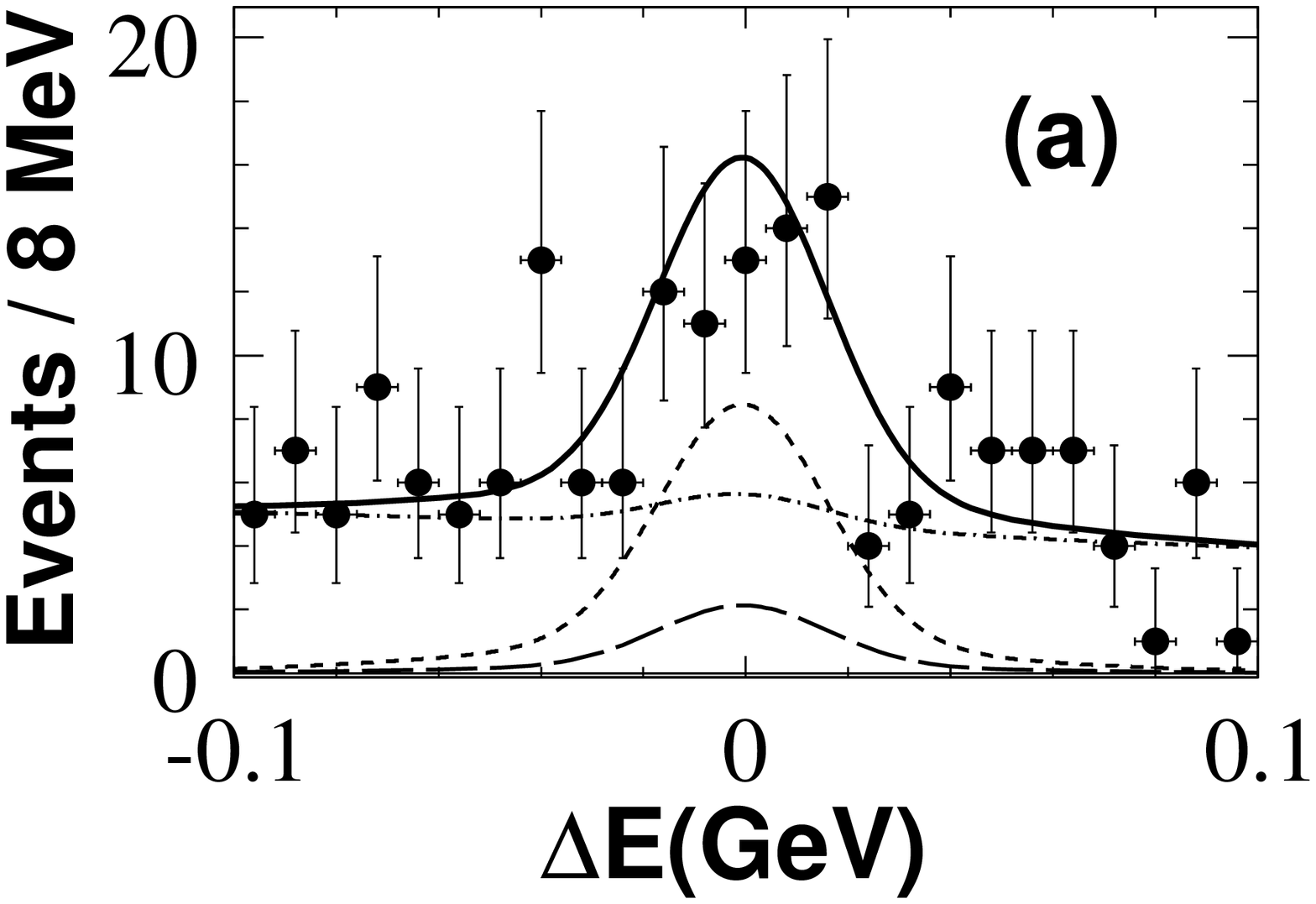}
~~~
\setlength{\epsfxsize}{0.35\linewidth}\leavevmode\epsfbox{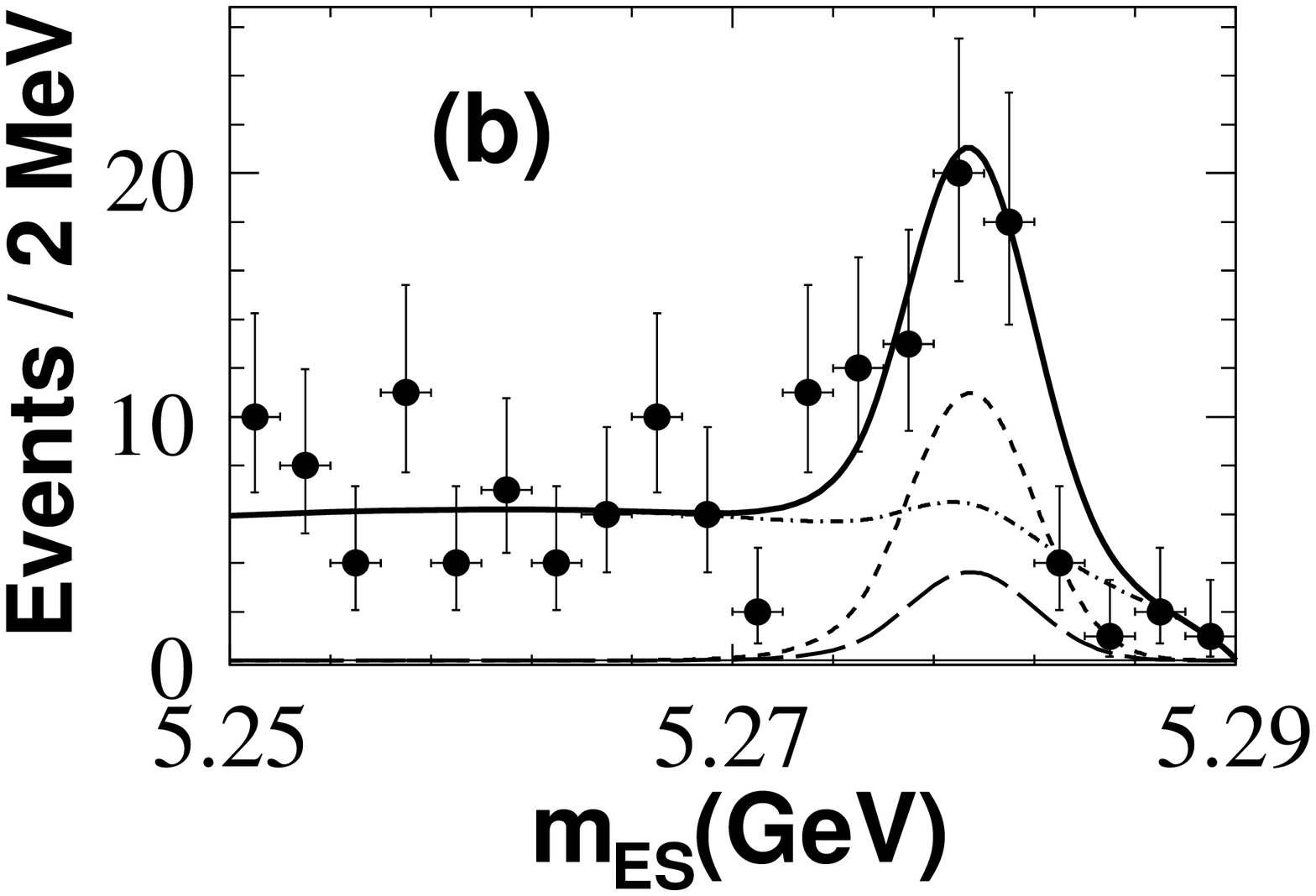}
}
\centerline{
\setlength{\epsfxsize}{0.35\linewidth}\leavevmode\epsfbox{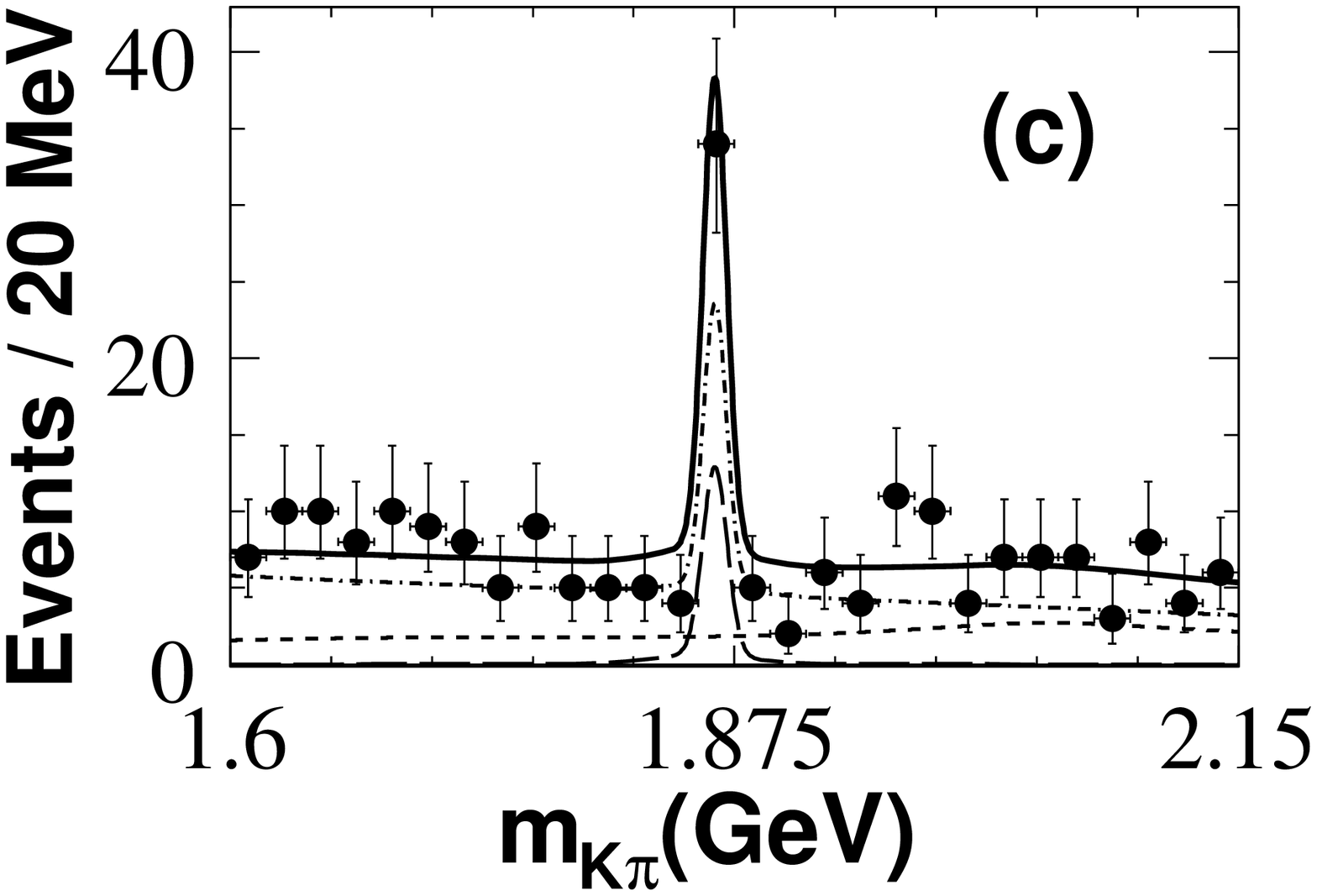}
~~~
\setlength{\epsfxsize}{0.35\linewidth}\leavevmode\epsfbox{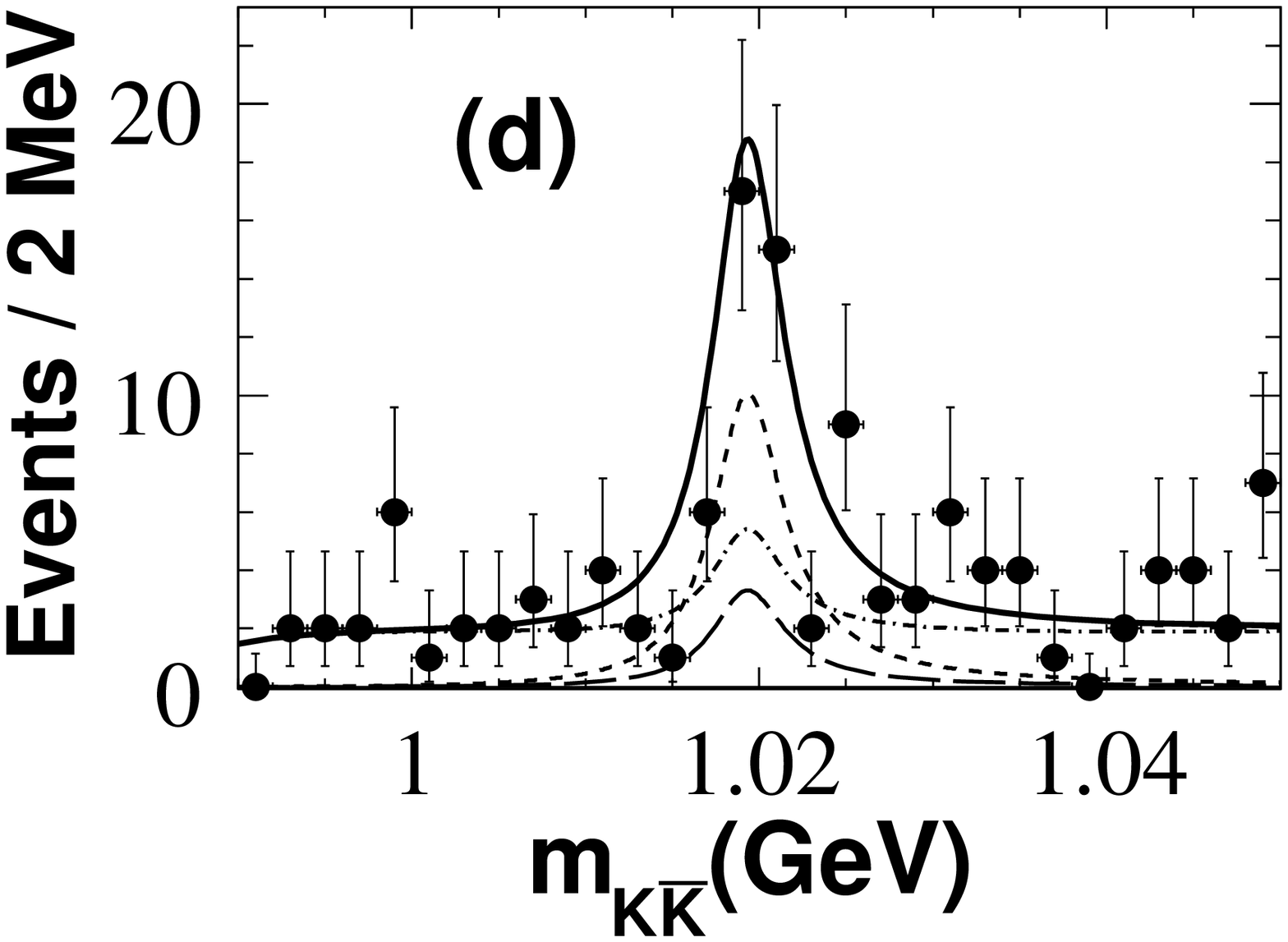}
}
\centerline{
\setlength{\epsfxsize}{0.35\linewidth}\leavevmode\epsfbox{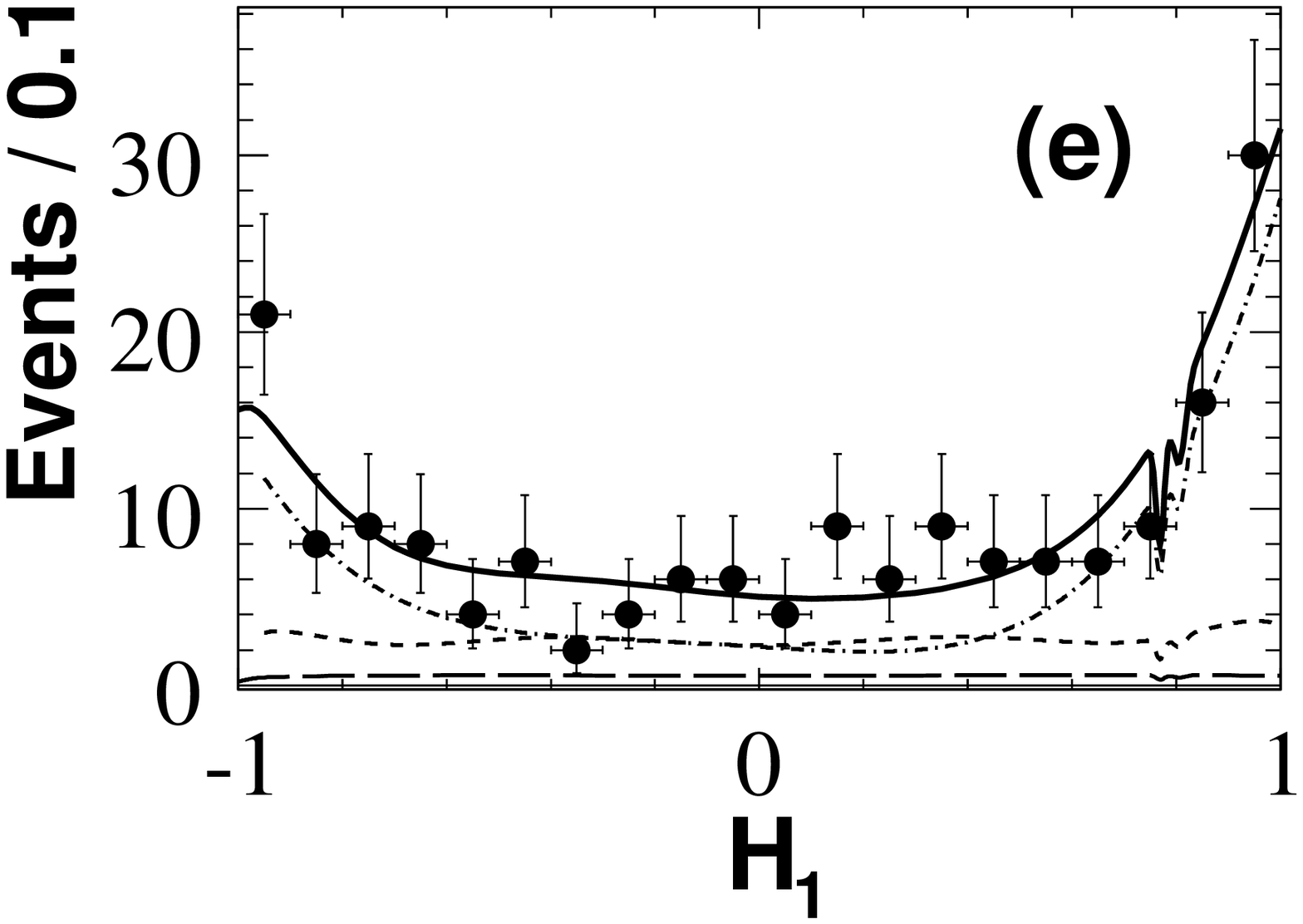}
~~~
\setlength{\epsfxsize}{0.35\linewidth}\leavevmode\epsfbox{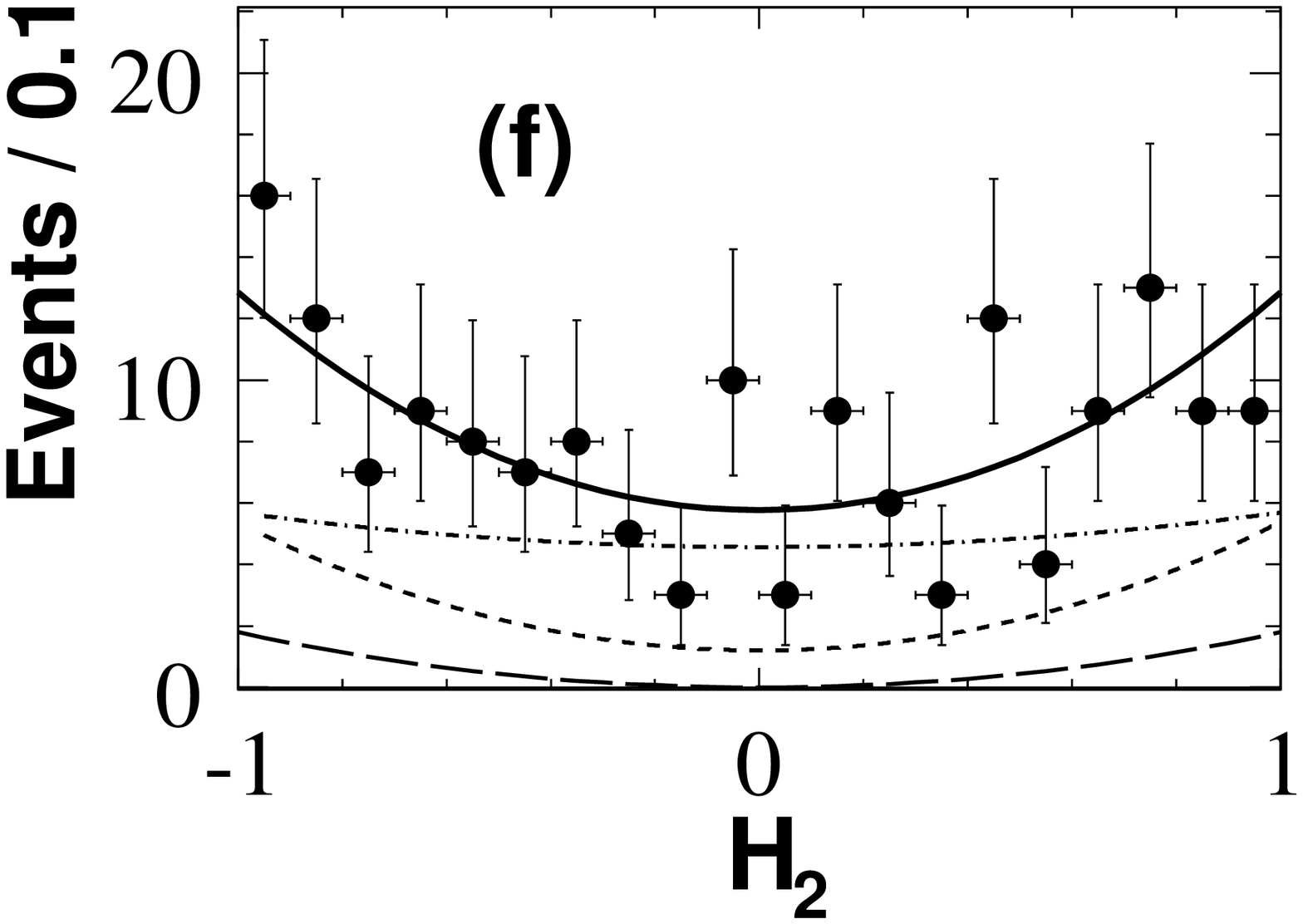}
}
\vspace{-0.3cm}
\caption{\label{fig:projection1} 
Projections onto the variables  $\Delta E$ (a), $m_{\rm ES}$ (b), 
$m_{K\!\pi}$ (c), $m_{K\!\Kbar}$ (d), ${\cal H}_1$ (e), and ${\cal H}_2$ (f)
for the signal $B^0\to\phi(K\pi)$ and $B^0\to\phi\Dbar^0$ candidates.
The solid (dotted-dashed) line shows the signal-plus-background 
(background only) PDF projection,
while the dashed (long-dashed) line 
shows PDF projection for the sum of four $B^0\to\phi(K\pi)$ 
categories (for $B^0\to\phi\Dbar^0$).
The pronounced $D^0$ mass peak in (c) is predominantly due to background.
The $D^+_{(s)}$-meson veto causes the sharp acceptance dips seen in (e).
}
\end{figure*}


We use two kinematic variables:
$\Delta{E}=(E_iE_B-\mathbf{p}_i\cdot\mathbf{p}_B-s/2)/\sqrt{s}$
and $m_{\rm{ES}} = [{ (s/2 + \mathbf{p}_i \cdot 
\mathbf{p}_B)^2 / E_i^2 - \mathbf{p}_B^{\,2} }]^{1/2}$,
where $(E_i,\mathbf{p}_i)$ is the 
$e^+e^-$ initial state four-momentum, 
and $(E_B,\mathbf{p}_B)$ is the four-momentum of the $B$ candidate.
We require $|\Delta{E}|<0.1$ GeV and $m_{\rm{ES}}>5.25$ GeV.
The requirements on the invariant masses are 
$1.60 < m_{K\!\pi} < 2.15$ GeV and
$0.99 < m_{K\!\Kbar} < 1.05$ GeV.

To reject the dominant $e^+e^-\to$ quark-antiquark continuum background, 
we use event-shape variables calculated in the center-of-mass frame.
We require $|\cos\theta_T| < 0.8$, where $\theta_T$ 
is the angle between the $B$-candidate thrust axis
and that of the rest of the event.
We construct a Fisher discriminant, ${\cal F}$,
that combines the polar angles of the $B$-momentum vector 
and the $B$-candidate thrust axis with respect to the beam axis,
and two moments of the energy flow around 
the $B$-candidate thrust axis~\cite{bigPRD}.

We remove signal candidates that have decay products 
with invariant mass within 12 MeV of the nominal mass 
values for $D_s^+$ or $D^+\to{\phi\pi^+}$.
In about $8.8\%$ of events more than one candidate is 
reconstructed and we select the one whose four-track 
vertex has the lowest $\chi^2$. 


We use an unbinned, extended maximum-likelihood fit~\cite{babar:vt} 
to extract the event yields $n_j$ and the probability density function 
(PDF) parameters, denoted by {\boldmath$\zeta$}$=\{f^1_L,f^3_L,f^4_L\}$
for the polarization parameters and {\boldmath$\xi$}
for the remaining parameters.
The data model has eight event categories $j$: 
$B^0\to\phi(K\pi)_0^{*0}$,
$\phi K^{*}(1680)^{0}$, 
$\phi K_3^{*}(1780)^{0}$, 
$\phi K_4^{*}(2045)^{0}$,
$\phi\Dbar^0$, 
$f_0(980)K^{*0}$, 
$f_0(980)\Dbar^0$,
and combinatorial background.
The $f_0(980)K^{*0}$ and $f_0(980)\Dbar^0$ categories
are included to account for both the resonant and 
nonresonant $K^+K^-$ contribution in exclusive $B$ decays, 
while the combinatorial background PDF is found to account 
well for both the dominant quark-antiquark background 
and the random tracks from the $B$ decays.

The likelihood ${\cal L}_i$ for each candidate $i$ is defined as
${\cal L}_i = \sum_{j}n_{j}\, 
{\cal P}_{j}$({\boldmath ${\rm x}_i$};~{\boldmath$\zeta$};~{\boldmath$\xi$}),
where each of the ${\cal P}_{j}$ is the PDF for variables
{\boldmath ${\rm x}_i$}$=\{\Delta E$, $m_{\rm{ES}}$, 
$m_{K\!\pi}$, $m_{K\!\Kbar}$, 
${\cal H}_1$, ${\cal H}_2$, ${\cal F}\}$.
We do not allow $C\!P$-violation in the decay amplitudes and ignore 
interference between the final states $B\to\phi(K\pi)_{J}$
with different $J$ because no significant signal is observed.
Since our acceptance in the decay angles is nearly uniform,
the event yields are almost completely unaffected by
interference among states of different $J$.

The PDF 
${\cal P}_{j}$({\boldmath ${\rm x}_i$};~{\boldmath$\zeta$};~{\boldmath$\xi$}) 
for a given candidate $i$ is a joint PDF for the helicity angles, 
and the product of the PDFs for each of the remaining variables.
The helicity part of the exclusive $B$ decay PDF is the 
ideal angular distribution from Eq.~(\ref{eq:helicityfull})
averaged over azimuthal angle $\Phi$,
where the amplitudes $A^J_{\lambda}$ are expressed in terms of
the polarization fractions {\boldmath$\zeta$}, multiplied 
by an empirically-determined acceptance function
${\cal{G}}({\cal H}_1,{\cal H}_2)
\equiv{\cal{G}}_1({\cal H}_1)\times{\cal{G}}_2({\cal H}_2)$.

A relativistic spin-$J$ Breit-Wigner amplitude
parameterization is used for the resonance 
mass~\cite{pdg2006,f0mass}, except for the nonresonant 
$(K\pi)^{*0}_0$ contribution which has no $m_{K\!\pi}$ 
amplitude dependence beyond the phase-space factor.
In the previous analysis with the ${K\pi}$ mass below 1.6 GeV, 
we parameterized the $(K\pi)^{*0}_0$ 
$m_{K\!\pi}$ amplitude with the LASS function~\cite{babar:vt,Aston:1987ir}, 
which includes the $K_0^{*}(1430)^0$ resonance together with 
a nonresonant component. However, above 1.6 GeV the validity of the
LASS parameterization is not certain and we use the phase-space model
for the nonresonant $(K\pi)^{*0}_0$ parameterization. 

The parameters {\boldmath$\xi$} describe the 
background or the remaining signal PDFs. 
They are left free to vary in the fit for the combinatorial 
background or are fixed to the values extracted from 
Monte Carlo (MC) simulation~\cite{geant} and calibration of 
$B$-decay channels for the exclusive $B$ decays. 
We use a sum of Gaussian functions 
for the parameterization of the signal PDFs 
for $\Delta E$, $m_{\rm{ES}}$, ${\cal F}$, 
and of the $D^0$ meson $m_{K\!\pi}$ distribution.
For the combinatorial background, we use polynomial functions,
except for $m_{\rm{ES}}$ and ${\cal F}$ distributions
which are parameterized by an empirical phase-space 
function~\cite{argus}
and by Gaussian functions, respectively.
The $\phi$ and $D^0$ meson production can occur in the 
background, and we take this into account in the PDF.


In the mass range $1.60 < m_{K\!\pi} < 2.15$ GeV,
we do not find significant signal 
in any of the four decays $B^0\to\phi(K^+\pi^-)$ with 
$K^{*}(1680)^{0}$, $K_3^{*}(1780)^{0}$, $K_4^{*}(2045)^{0}$, or 
$\Dbar^0\to K^+\pi^-$
and we place limits on their branching fractions
as shown in Table~\ref{tab:results1}.
We see evidence for the nonresonant $\phi (K\pi)^{*0}_0$
contribution consistent with extrapolation (33 events) 
from the lower mass range studied in~Ref.~\cite{babar:vt}. 
Due to large correlation among various signal yields
of the decay modes with broad $K\pi$ distributions,
the errors on individual decay modes are relatively large.
However, the significance of the $B^0\to\phi(K^+\pi^-)$
decay with $(K\pi)^{*0}_0$, $K^{*}(1680)^{0}$,
$K_3^{*}(1780)^{0}$, and $K_4^{*}(2045)^{0}$ combined
is larger than 5$\sigma$.
The significance is defined as the square root of the change in 
$2\ln{\cal L}$ when the yield is constrained to zero in the 
likelihood ${\cal L}$.

Since we do not determine the flavor 
of the neutral $B$ meson, our limits refer to the sum of two 
flavor final states, such as $\phi\Dbar^{0}$ and $\phi D^{0}$.
We assume equal production of $B^+B^-$ and $B^0\Bbar^0$ pairs
in $\Upsilon(4S)$ decays. 
In Fig.~\ref{fig:projection1} we show projections onto the variables.
Data distributions are shown with a requirement on the signal-to-background
probability ratio calculated with the plotted variable excluded.
This requirement is optimized to enhance the signal
and results in signal selection efficiency between 60\% and 90\%. 

In the fit, we constrain both event yields and polarization
fractions $f_L^J$ to the physically allowed ranges. 
The negative event yield in the $B^0\to\phi K_3^{*}(1780)^{0}$
decay is obtained by using the likelihood in the positive event region
and fitting its shape with a parabolic function whose mininum is in the 
negative event region.
For the three $B^0\to\phi K_J^{*0}$ decay modes 
with $J\ge1$, the $f_L^J$ fit results are consistent 
with any allowed value between 0 and 1 and we assume polarization 
which gives the smallest reconstruction
efficiency in the branching fraction calculation.
We integrate the likelihood distributions in the physically allowed
ranges to compute the upper limits on the branching fractions.


The nonresonant $K^+K^-$ contribution under the 
$\phi$ is accounted for with the $B^0\to f_0 K^{*0}$ category
with the broad $f_0$ invariant mass distribution~\cite{f0mass}.
Its yield is consistent with zero for any of the $K^{*0}$ spin assumptions.
We find evidence for a nonzero event yield in this nonresonant
$K^+K^-$ region under the $\phi$ with a $\Dbar^0$ of $(31^{+9}_{-8})$ events,
with statistical errors only quoted. However, due to the broad $K^+K^-$ 
invariant mass distribution, we cannot distinguish between 
$f_0$, $a_0$, or any other broad $K^+K^-$ contribution under the $\phi$.
The uncertainties due to $m_{K\!\Kbar}$ parameterization are
estimated with variation of the shape model from the resonant $f_0$ 
to phase-space and account for the errors between 3 and 11 events
in different channels.

We vary those parameters in {\boldmath$\xi$} not used 
to model combinatorial background within their 
uncertainties and derive the associated systematic errors
between one and three events.
The signal PDF model excludes the fake combinations originating
from misreconstructed events.
The biases from the dilution due to the presence of fake combinations, 
the finite resolution of the angle measurement, 
or other imperfections in the signal PDF model are estimated 
with MC simulation and generated samples. This results 
in an uncertainty between 1 and 11 events. 

Additional systematic uncertainty originates 
from $B$ background, where we estimate that only a few events 
can fake the signal.
The systematic errors in selection efficiencies are dominated 
by those in particle identification (4\%), track finding (2\%),
and uncertainty due to the $K^*$ resonance parameters~\cite{pdg2006}
of 2--13\%.
Other systematic effects arise from event-selection criteria, 
$\phi$, $K^{*0}$, or $D^0$ branching fractions~\cite{pdg2006}, 
and number of $B$ mesons.


Our results place stringent limits on the $B^0\to\phi K^{*0}$
branching fractions with the higher-mass and spin
resonances $K^{*}(1680)^{0}$, $K_3^{*}(1780)^{0}$, 
and $K_4^{*}(2045)^{0}$ when compared with 
the lower-mass states~\cite{babar:vv,belle:phikst,babar:vt}.
The decay rate suppression may serve as an additional
tool to study the mechanism of the anomalous
transverse amplitude in the $B\to\phi K^{*}(892)$ decay.
We find the $B^0\to\phi(K\pi)_0^{*0}$ rate with scalar
$(K\pi)_0^{*0}$ to be consistent for $K\pi$ invariant mass
above and below 1.6 GeV. 
Our limit on the $B^0\to\phi\Dbar^0$ decay provides a test
of the $B$ decay mechanisms involving $\phi$ mesons 
in the final state.

In summary, we have searched for the
$B^0\to\phi K^{*0}$ decays with the tensor
$K_3^{*}(1780)^{0}$ and $K_4^{*}(2045)^{0}$, 
vector $K^{*}(1680)^{0}$, and scalar nonresonant $(K\pi)_0^{*0}$
contributions
with $K^{*0}\to K^+\pi^-$ invariant mass above 1.6 GeV.
Our results are summarized in Table~\ref{tab:results1}. 
We do not find significant signal with the above resonances and
place upper limits on these and $B^0\to\phi\Dbar^0$ decays.


We are grateful for the excellent luminosity and machine conditions
provided by our \pep2\ colleagues,
and for the substantial dedicated effort from
the computing organizations that support \babar.
The collaborating institutions wish to thank
SLAC for its support and kind hospitality.
This work is supported by
DOE
and NSF (USA),
NSERC (Canada),
IHEP (China),
CEA and
CNRS-IN2P3
(France),
BMBF and DFG
(Germany),
INFN (Italy),
FOM (The Netherlands),
NFR (Norway),
MIST (Russia),
MEC (Spain), and
PPARC (United Kingdom).
Individuals have received support from the
Marie Curie EIF (European Union) and
the A.~P.~Sloan Foundation.


\bibliographystyle{h-physrev2-original}   

\end{document}